\documentclass[12pt,jpa]{iopart}

\usepackage[numbers,sort&compress]{natbib}
\usepackage{rotating}
\usepackage{graphicx}
\usepackage{color}
\usepackage{caption}

\newcommand\eq[1]{Eq.~(\ref{#1})}
\newcommand\reff[1]{(\ref{#1})}

\newcommand{\sign}{\textup{sign}}

\begin{document}
	
\title[Casimir Force in a Model with Strongly Adsorbing Competing Walls]{Casimir Force in a Model with Strongly Adsorbing Competing Walls:  Analytical Results}
	

\author{Daniel M Dantchev, Vassil M Vassilev and Peter A Djondjorov}
\address{Institute of Mechanics -- Bulgarian Academy of Sciences, \\
	 Acad. Georgy Bonchev St., Building 4,	1113 Sofia, Bulgaria}
\ead{daniel@imbm.bas.bg, vasilvas@imbm.bas.bg and padjon@imbm.bas.bg}


\begin{abstract}
	We present both analytical and numerical results for the behaviour of  the Casimir force in a Ginzburg-Landau type model of a film of a simple fluid or binary liquid mixture in which the confining surfaces are strongly adsorbing but preferring different phases of the simple fluid, or different components of the mixture. Under such boundary conditions an interface is formed between the competing phases inside the system which are forced to coexist. We investigate the force as a function of the temperature and in the presence of an external ordering field and determine the (temperature-field) relief map of the force. We prove the existence of a single global maximum of the force and find its position and value. We find the asymptotic behavior of the force when any of the scaling fields becomes large while the other one is negligible. Contrary to the case of symmetric boundary conditions we find, as expected, that the finite system does not possess a phase transition of its own for any finite values of the scaling variables corresponding to the temperature and the ordering field. We perform the study near the bulk critical temperature of the corresponding bulk system and find a perfect agreement with the finite-size scaling theory. 
\end{abstract}


\section{INTRODUCTION}

In the current article we study the behaviour of a system with  $\infty^2 \times L$ film geometry.  Let us recall that  when at least one of the spacial extensions of the system is finite, one terms the corresponding system a finite one. In a system with a film geometry the pressure acting on its bounding surfaces differs, generally speaking, from the pressure that will be acting in the same system, characterized by the same thermodynamic parameters, but being infinite in extend. In different branches of science and in different range of parameters one uses the terms disjoining pressure, solvation force, or the thermodynamic Casimir force \cite{E90book, K94,  BDT2000}.   The Casimir like forces exist in any system where massless excitations are spatially limited by the presence of material bodies. The term “thermodynamic Casimir” force encompasses  two cases with such massless excitations existing in thermodynamic (statistical mechanical) systems: near $T_c$, where they are the critical fluctuations of the order parameter and one speaks about the {\em critical Casimir force}, and below $T_c$, in systems with Goldstone modes, say $^4$He, liquid crystals,  where one sometimes speaks about off-critical Casimir force. 

Let us consider a system with a film geometry at a temperature $T$ exposed to an external ordering field $h$ that couples to its order parameter - density, concentration difference, magnetization, etc., and let $(T=T_c, h=0)$ is its bulk critical point in the $(T,h)$ plane. The thermodynamic Casimir force $F_{\rm Cas}(T,h,L)$ in such a system is the excess pressure over the bulk one acting on the boundaries of the finite system which is due to the finite size of that system, i.e., 
\begin{equation} \label{Casimir}
F_{\rm Cas}(T,h,L)= P_L(T,h)-P_b(T,h).
\end{equation} 
Here $P_L$ is the pressure in the finite system, while $P_b$ is that one in the infinite system. Let us note that the above definition is actually equivalent to another one which is also commonly used \cite{E90book,K94,BDT2000}
\begin{equation}
\label{grand_can}
F_{\rm Cas}(T,h,L)\equiv-\frac{\partial\omega_{\rm ex}(T,h,L)}{\partial L}=-\frac{\partial\omega_L(T,h,L)}{\partial L}-P_b,
\end{equation}
where $\omega_{\rm ex}=\omega_L-L\,\omega_b$ is the excess grand potential per unit area, $\omega_L$ is the grand canonical potential of the finite system, again per unit area, and $\omega_b$ is the density of the grand potential for the infinite system. The equivalence between the definitions \eq{Casimir} and \eq{grand_can} comes from the observation that  $\omega_b=- P_b$ and for the finite system with surface area $A$ and thickness $L$ one has $\omega_L=\lim_{A\to \infty} \Omega_L/A$, with $-\partial\omega_L(T,h,L)/\partial L=P_L$. 

When a finite system is thermodynamically positioned close to its critical temperature its two-points correlation length $\xi$ becomes comparable to  $L$, and then the thermodynamic functions describing its behaviour depend on the dimensionless ratio $L/\xi$ taking a scaling form given by the finite-size scaling theory \cite{Ba83,C88,Ped90,PE90,BDT2000}.

When a {\it finite} system undergoes a phase transition its phase behaviour can be significantly different from that in the bulk counterpart  \cite{FB72,Ba83,C88,Bb83,D86,Di88,P90,E90book,BDT2000}. One observes effects like shifts of the bulk critical points, that can be both with respect to $T$ and $h$, or the appearance of phase transitions of its own at some new critical point $(T_{c,L}, h_L)$, like the capillary condensation phase transition, if the dimensionality of the finite system is large enough. Near the bulk critical point, the behaviour of the bulk system is characterized with critical exponents and scaling functions which depend only on gross features of the system like the dimensionality of the system $d$, the symmetry of the ordered state, normally denoted by $n$, and the long-ranginess of the interaction involved, i.e., on the so-called bulk universality class. The behaviour of the finite system, in addition, depends on the so-called surface universality class, which is determined  by the boundary conditions on the surfaces of the finite system, as well as on its geometry. In a system with a film geometry if the finite system exhibits a phase transition of its own it belongs to the universality class of the $(d-1)$-dimensional infinite system. One of the quantities of especial interest in a finite critical system that is a subject of a plethora of studies, is the thermodynamic Casimir force and, more especially, that one which is observed near the critical point of the bulk system, often termed, as already explained above, {\it critical Casimir force}. 

The critical Casimir effect has been already directly observed, utilizing light scattering measurements, in the interaction of a colloid spherical particle with a plate \cite{HHGDB2008} both of which are immersed in a critical binary liquid mixture. Very recently the nonadditivity of critical Casimir forces has been experimentally demonstrated in \cite{PCTBDGV2016}. Indirectly, as  a balancing force that determines the thickness of a wetting film in the vicinity of its bulk critical point the Casimir force has been also studied in $^4$He \cite{GC99}, \cite{GSGC2006}, as well as in $^3$He--$^4$He mixtures \cite{GC2002}. In  \cite{FYP2005} and   \cite{RBM2007} measurements of the Casimir force in thin wetting films of binary liquid mixture are also performed. The studies in the field have also enjoined a considerable theoretical attention.  Reviews on the corresponding results can be found in \cite{K99,G2009,TD2010,GD2011,D2012,V2015}. 

Before turning exclusively to the behaviour of the Casimir force, let us briefly remind some basic facts of the theory of critical phenomena. In the vicinity of the bulk critical point  $(T_c,h=0)$ the bulk correlation length of the order parameter $\xi$ becomes large, and theoretically diverges: $\xi_t^+\equiv\xi(T\to T_c^{+},h=0)\simeq \xi_0^{+} t^{-\nu}$, $t=(T-T_c)/T_c$, and $\xi_h\equiv\xi(T=T_c,h\to 0)\simeq \xi_{0,h} |h/(k_B T_c)|^{-\nu/\Delta}$, where $\nu$ and $\Delta$ are the usual critical exponents and $\xi_0^{+}$ and $\xi_{0,h}$ are the corresponding nonuniversal amplitudes of the correlation length along the $t$ and $h$ axes.  For temperatures such that the correlation length  $\xi$ becomes comparable to  $L$, the thermodynamic functions describing its behaviour depend on the ratio $L/\xi$ and take scaling forms given by the finite-size scaling theory. For such a system the finite-size scaling theory \cite{C88,BDT2000,Ba83,P90,Ped90,K94} predicts for the Casimir force
\begin{equation}\label{cas}
F_{\rm Cas}(t,h,L)=L^{-d}X_{\rm Cas}(x_t,x_h);
\end{equation}
where
$x_t=a_t t L^{1/\nu}$, $x_h=a_h h L^{\Delta/\nu}$.
Here $d$ is the dimension of the system, $a_t$ and $a_h$ are nonuniversal metric factors that can be fixed, for a given system, by taking them to be, e.g., $a_t=1/\left[\xi_0^+\right]^{1/\nu}$, and $a_h=1/\left[\xi_{0,h}\right]^{\Delta/\nu}$. 

In the next section we are going to consider the Casimir force within the  Ginzburg-Landau mean-field model. Let us recall that the model we are going to consider is a standard model within which one studies phenomena like critical adsorption \cite{PL83,E90,FD95,THD2008,BU2001,EMT86,OO2012,MCS98,DSD2007,DMBD2009,C77,G85,TD2010,DRB2007,DVD2015},  wetting or drying \cite{C77,NF82,G85,BME87,Di88,SOI91}, surface phenomena \cite{Bb83,D86}, capillary condensation \cite{BME87,E90,BLM2002,EMT86,OO2012,DSD2007,YOO2013,DVD2015}, localization-delocalization phase transition \cite{PE90,PE92,BLM2003}, finite-size behaviour of thin films \cite{KO72,NF83,FN81,BLM2003,NAFI83,EMT86,Ba83,C88,Ped90,PE90,BDT2000,DVD2015,DVD2016}, the thermodynamic Casimir effect \cite{INW86,K97,SHD2003,GaD2006,DSD2007,MGD2007,ZSRKC2007,DVD2016}, etc.  The results of the model have been also used to calculate the Casimir forces in systems with chemically or topographically patterned substrates, as well as, coupled with the Derjaguin approximation, for studies on interactions of colloids -- see, e.g., the review \cite{GD2011} and the literature cited therein. Until very recently, i.e. before Ref. \cite{DVD2015,DVD2016}, the results 
for the case $h=0$ were derived analytically \cite{INW86,K97,GaD2006,DRB2009} while the $h$-dependence was studied numerically either at the bulk critical point of the system $T=T_c$, or along some specific isotherms -- see, e.g., \cite{SHD2003,MB2005,DSD2007,DRB2007,DRB2009,PE92,LTHD2014}. In the current article we are going to improve this situation with respect to the Casimir force by deriving exact analytical results for it in the $(T,h)$ plane for the case of a film system bounded by strongly adsorbing {\it competing} walls. These studies compliments our recent results for the same model in the case when the walls equally prefer the same phase of the bounded fluid \cite{DVD2015,DVD2016}. Let us note that in the case of competing walls, quite surprisingly, and contrary to the symmetric case, the finite system does not posses a phase transition of its own below some $T_{c,L}$ that is algebraically close to $T_c$ on the scale of $L^{-1/\nu}$. This has been an object of series of studies - see, e.g.,  \cite{PEN91,PE90,BLF95,BELF96,BLF95b,BLM2003,BHVV2008,AB2009,VD2013}. In this specific case a phase transition of its own turns out to occur for $T_{c,L}<T_w$, where $T_w$ is the wetting temperature of the corresponding semi-infinite system. Let us note that, as a rule, $T_{w}$ is at a finite distance below $T_c$. 

Before going into specific calculations, let us note that one normally models the preference of a wall to a given phase or component of the binary mixture by imposing a surface fields $h_1$ and $h_L$ acting locally on the boundaries.  When the system undergoes
a phase transition in its bulk in the presence of such surface ordering fields one
speaks about the "normal" transition \cite{BD94}. It has been shown that it is equivalent, as
far as the leading critical behavior is concerned, to the "extraordinary" transition \cite{BD94,D86} which is achieved by enhancing the surface couplings stronger than the bulk couplings.
It has been
demonstrated \cite{D86} that when $h_1 h_L\ne 0$ for the {\it leading} critical behavior of the system is
sufficient to investigate the limits $h_1, h_L\to \pm \infty$. Since we are interested in the leading finite-size behavior of a system with competing walls near its bulk critical point, in the current article we focus on the case of $h_1\to +\infty$, while $h_L\to -\infty$. One
usually refers to this case as the (+, -) boundary conditions. Thus, in the current article we will be only dealing with the behavior of the system under (+, -) boundary conditions.

\section{THE GINZBURG-LANDAU MEAN-FIELD MODEL AND THE CASIMIR FORCE}

\subsection{Definition of the Model}
We consider a critical system of Ising type in a film geometry $\infty^2\times L$, where $L$ is supposed to be along $z$ axis, described by the minimizers of the standard $\phi^4$ Ginzburg-Landau functional
\begin{equation} \label{FviafIsing}
{\cal F}\left[\phi;\tau,h,L\right] =\int_{-L/2}^{L/2} {\cal L}(\phi,\phi') dz,
\end{equation}
where
\begin{equation}\label{fdefIsing}
{\cal L}\equiv {\cal L}(\phi,\phi')=\frac{1}{2}  {\phi'}^2 +
\frac{1}{2}\tau\phi^2+\frac{1}{4}g\phi^4-h \phi.
\end{equation}
Here $L$ is the film thickness, ${\phi}(z|\tau,h,L)$ is the order parameter   assumed to depend on the perpendicular position $z\in(-L/2,L/2)$ only, $\tau=(T-T_c)/T_c\,(\xi_0^{+})^{-2}$ is the bare reduced temperature, $h$ is the external ordering field, $g$ is the bare coupling constant and the primes indicate differentiation with respect to the variable $z$.

\subsection{Basic expression for the Casimir force}

In Ref. \cite{DVD2016} we have shown that for the model  considered here the pressure in the finite system is
\begin{equation}
\label{dz}
P_L(\tau,h) = \frac{1}{2}  {\phi'}^2-\frac{1}{4}g\phi^4 - 
\frac{1}{2}\tau\phi^2+h \phi, 
\end{equation}
while in the bulk system it equals to
\begin{equation}\label{pb}
P_b(\tau,h)=-\frac{1}{4}g \phi_b^4-\frac{1}{2}\tau \phi_b^2+h \phi_b.
\end{equation}
Here 

{\it i)} The variable $\phi_b$ is the order parameter of the bulk system. It is determined by the cubic equation
$-\phi_b\left[\tau+g\,\phi_b^2\right]+h=0$, $\phi_b$ being such that
${\cal L}_b =
\frac{1}{2}\tau\phi_{b}^2+\frac{1}{4}g\phi_{b}^4-h \phi_{b}$
attains its minimum. Let us note that $P_b=-{\cal L}_b$, i.e., the $P_b$ has its {\it maximum} over the possible solutions of the cubic equation for $\phi_b$.  

{\it ii)} The order parameter of the finite system $\phi$ minimizes the functional ${\cal F}$. It is determined by the solutions of the corresponding Euler-Lagrange equation
\begin{equation}\label{Lagrange}
\frac{d}{dz} \frac{\partial {\cal L}}{\partial \phi'}-\frac{\partial {\cal L}}{\partial\phi}=0,
\end{equation}
which, on account of Eq. (\ref{fdefIsing}), reads
\begin{equation}\label{Phieqalone}
{\phi''}-\phi\left[\tau+g\,\phi^2\right]+h=0.
\end{equation}
Multiplying the above equation by $\phi'$ and integrating over $z$ one obtains that $P_L$ in \eq{dz} is, actually, the first integral of the system. 

When $F_{\rm Cas}(\tau,h,L)<0$ the excess pressure will be inward
of the system that corresponds to an {\it attraction} of the surfaces of the system towards each other and to a {\it repulsion} if $F_{\rm Cas}(\tau,h,L)>0$. 

Let us note that Eqs. (\ref{Casimir}) -- (\ref{pb}) does not depend on the boundary conditions applied on the surfaces of the system. The dependence on them enters, of course, via the dependence of the order parameter profile on the corresponding boundary conditions. Next, let us also stress that the approach outlined above is in fact equivalent to the one introduced in \cite{PE92}, see Section 2.1 there, where a more general form of the functional \eq{FviafIsing} has been used that allows for finite surface fields and surface enhancements. Then the model enjoys a rich phase diagram that has been obtained in \cite{SOI91} and \cite{PE92}. The model we are studying in the current article can be considered as a particular limiting case of that more general model with the surface fields amplitudes approaching plus infinity on one surface and minus infinity on the other. Let us note that under such a limit the wetting temperature of the corresponding semiinfinite system $T_w$ diminishes. 
 
In Ref. \cite{DVD2016} we have determined the behaviour of the Casimir force as a function of both the temperature and the external ordering fields for the case of the so-called $(+,+)$ boundary conditions for which $\left. \lim \phi\left( z \right) \right| _{z \rightarrow -L/2}=\left.
\lim \phi \left( z \right) \right| _{z \rightarrow L/2}=+\infty $.  In order to do that a detailed knowledge of the behaviour of the order parameter profile $\phi$ in the finite system has been needed. The behaviour of $\phi$ as a function of $\tau$ and $h$, as well as the corresponding response functions have been studied in \cite{DVD2015}. In the current article we will study the behaviour of the Casimir force for the case of the so-called $(+,-)$ boundary conditions
\begin{equation}
\label{eq:pl_min_bc}
\left. \lim \phi\left( z \right) \right| _{z \rightarrow -L/2}=+\infty, \left.
\lim \phi \left( z \right) \right| _{z \rightarrow L/2}=-\infty.
\end{equation}
We will present a solution of the problem that does not rely on the detailed knowledge of the order parameter profile. Before proceeding with the details let us note that, as in the case of $(+,+)$ boundary conditions, the exact behaviour of the Casimir force under $(+,-)$ boundary conditions is known only for the $h=0$ case due to the results reported in \cite{K97}. Here, for the $(+,-)$ case we will extend these results for the case of a nonzero external field.   We are aware on only data for the behaviour of the force as a function of $h$ at $T=T_c$ reported in \cite{VD2013}, where they have been obtained numerically. We will demonstrate that there is an excellent agreement between the analytical results we are going to present and the numerical data in  \cite{VD2013}. 

\section{EXACT RESULTS FOR THE CASIMIR FORCE} 
Since the thermodynamic Casimir force is normally presented in terms of the scaling variables 
\begin{equation}\label{isvarxt}
l_t\equiv \sign(\tau) \, L/\xi_t^+ =\sign(\tau)\, L \sqrt{|\tau|},
\end{equation}
\begin{equation}\label{isvarxh}
l_h\equiv \sign(h)\, L/\xi_h=L\sqrt{3}\left(\sqrt{g}\, h\right)^{1/3}, 
\end{equation}
in the remainder we are going to use such variables as the basic parameters determining the behaviour of the force. In the above we  have taken into account that for the model considered here $\xi_{0,h}/\xi_0^+=1/\sqrt{3}$ \cite{SHD2003}, $\nu=1/2$ and $\Delta=3/2$.

\subsection{Exact results for the case of zero ordering field}

As stated already above, the corresponding result has been derived in \cite{K97}. There it has been shown that the scaling function of the Casimir force under $(+,-)$ boundary condition in the Ising mean-field model is
\begin{eqnarray}\label{CasimiralphaeqpiKrech}
\fl \lefteqn{X_{\rm Cas}^{(+,-)}(l_t, l_h=0)}\nonumber\\
\fl &=&\left\{ \begin{array}{ccc}
64\; k^2 (1-k^2)\left[K(k)\right]^4, & l_t=-[2\left[2K(k)\right]^2(2k^2-1)]^{1/2}, & l_t \le 0 \\
\left[2K(k)\right]^4,& l_t=[-2\left[2K(k)\right]^2(2k^2-1)]^{1/2}, & 0 \le l_t \le \sqrt{2}\pi \\
\left[2K(k)\right]^4(1-k^2)^2, & l_t=[2\left[2K(k)\right]^2(k^2+1)]^{1/2}, & l_t \ge \sqrt{2}\pi
\end{array},
\right.
\end{eqnarray}
where  $K$ is the complete elliptic integral of the first kind, with $k$ being its elliptic modulus.   It is interesting to note that
the scaling function $X_{\rm Cas}^{(+,-)}(l_t,0)$ is related to   $X_{\rm Cas}^{(+,+)}(l_t,0)$ through \cite{VGMD2009,BDR2011}
\begin{equation}\label{relscfunctionspppm}
X_{\rm Cas}^{(+,+)}(l_t,0)=-\frac{1}{4} X_{\rm Cas}^{(+,-)}(-l_t/2,0).
\end{equation}
The last, of course, implies that for the corresponding mean-field Casimir amplitudes one has
\begin{equation}\label{deltarelpppm}
\frac{\Delta_{\rm Cas}^{(+,+)}}{ \Delta_{\rm Cas}^{(+,-)}}=-\frac{1}{4}.
\end{equation}
We recall that for $l_h=0$ one has \cite{K97,BDR2011}
\begin{equation}
\label{Casimiralphaeq0Krech}
\fl X_{\rm Cas}^{(+,+)}(l_t,0)=\left\{ \begin{array}{ccc}
-4\left(1-k^2\right)^2K^4\left(k\right), & l_t=-[4(1+k^2)K^2(k)]^{1/2}, & l_t \le -\pi \\
-4 K^4\left(k\right),&l_t=-[4(1-2k^2)K^2(k)]^{1/2}, & 0 \ge l_t \ge -\pi\\
-4 k^2 (1-k^2) K^4\left(k\right), & l_t=[4(2k^2-1)K^2(k)]^{1/2}, & l_t \ge 0
\end{array}.
\right.
\end{equation}

\subsection{Exact results for the case of nonzero ordering field}

In terms of the scaling  variables $l_t$ and $l_h$ the value $P_L(\tau,h)$ of the first integral, see \eq{dz}, becomes 
\begin{equation} \label{FIX}
P_L(\tau,h)=\frac{1}{g L^4} p\left( l_t,l_h \right),
\end{equation}
where the constant $p\left( l_t,l_h \right)$ is 
\begin{equation} \label{eq:p_scaling}
p\left( l_t,l_h \right)={X'}^2
-X^4-\sign(l_t) \, l_t^2 X^2 +\frac{2}{3 \sqrt{6}}l_h^{3} X.
\end{equation}
Here
\begin{equation}\label{eq:pl_lt_lh}
X(\zeta|l_t,l_h)=\sqrt{\frac{g}{2}}L^{\beta/\nu} \phi(z|\tau, h, L) 
\end{equation}
is the scaling function of the order parameter $\phi$,  $\beta=1/2$ and hereafter the prime means differentiation with respect to the variable $\zeta=z/L, \zeta \in [-1/2,1/2]$. Similarly, for the bulk system, see \eq{pb}, one has 
\begin{equation} \label{pbX}
P_b(\tau,h)=  \frac{1}{g L^4}p_b(l_t,l_h),
\end{equation}
where
\begin{equation} \label{pbX1}
p_b(l_t,l_h)= -X_{b}^4-\sign(l_t) \, l_t^2 X_{b}^2 +\frac{2}{3 \sqrt{6}}l_h^{3} X_{b}.
\end{equation}
From Eqs. (\ref{FIX}) and (\ref{pbX}) for the Casimir force, see Eq. (\ref{Casimir}), one obtains 
\begin{equation} \label{CasimirX}
F_{\rm Cas}(\tau,h,L)= \frac{1}{g L^4} X_{\rm Cas}(l_t,l_h),
\end{equation}
where  $X_{\rm Cas}$ is its scaling function
\begin{equation} \label{CasimirX1}
X_{\rm Cas}(l_t,l_h)= p\left( l_t,l_h \right)-p_b(l_t,l_h).
\end{equation}

From \eq{eq:p_scaling} one can write the equation for the scaling function of the order parameter profile in the form 
\begin{equation} \label{eq:oder_parameter}
[X'(\zeta)]^2=P[X(\zeta)|l_t, l_h], 
\end{equation}
where
\begin{equation} \label{eq:oder_parameter_2}
P[X(\zeta)|l_t, l_h]=X^4+\sign(l_t) \, l_t^2 X^2 -\frac{2}{3 \sqrt{6}}l_h^{3} X+p\left( l_t,l_h \right).
\end{equation}
Due to the boundary conditions given in \eq{eq:pl_min_bc} the order parameter profile decays from $+\infty$ on the left boundary to $-\infty$ on the right plate, i.e., $X'(\zeta) \leq 0$ for $\zeta\in(-1/2,1/2)$. Thus, the equation for the order parameter profile implies 
\begin{equation}
\label{eq:or_par}
\frac{d X}{\sqrt{P[X]}}=-d\zeta. 
\end{equation}
Integrating this equation one obtains
\begin{equation}
\label{eq:main_eq}
1=f(l_t,l_h,p), 
\end{equation}
where
\begin{equation}
\label{eq:main_eq_2}
f(l_t,l_h,p)= \int_{-\infty}^{\infty} \frac{dX}{\sqrt{X^4+\sign(l_t) \, l_t^2 X^2 -\frac{2}{3 \sqrt{6}}l_h^{3} X+p\left(l_t,l_h \right)}}.
\end{equation}
\begin{figure}[h!]
	\includegraphics[width=8cm]{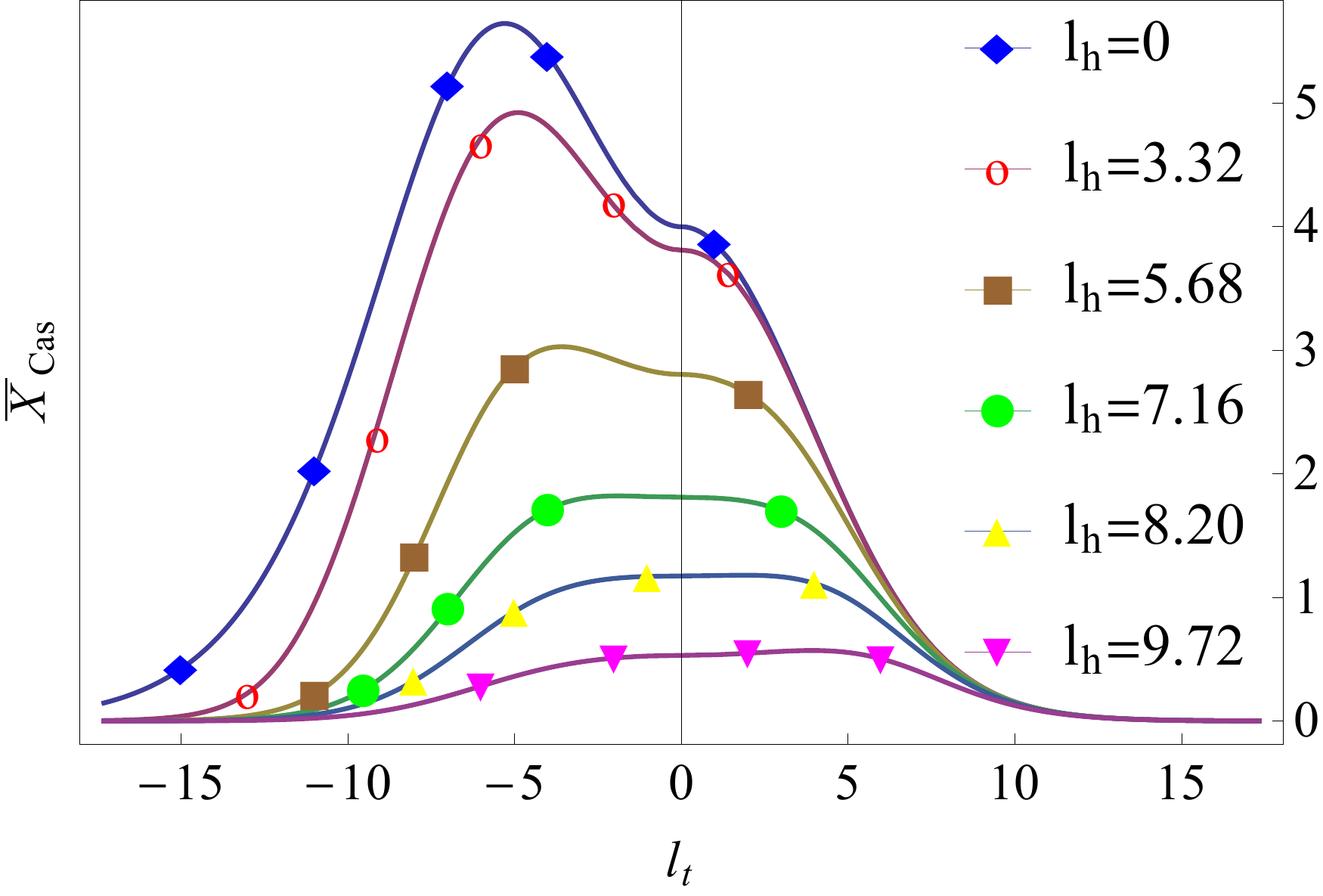}
	\includegraphics[width=8.4cm]{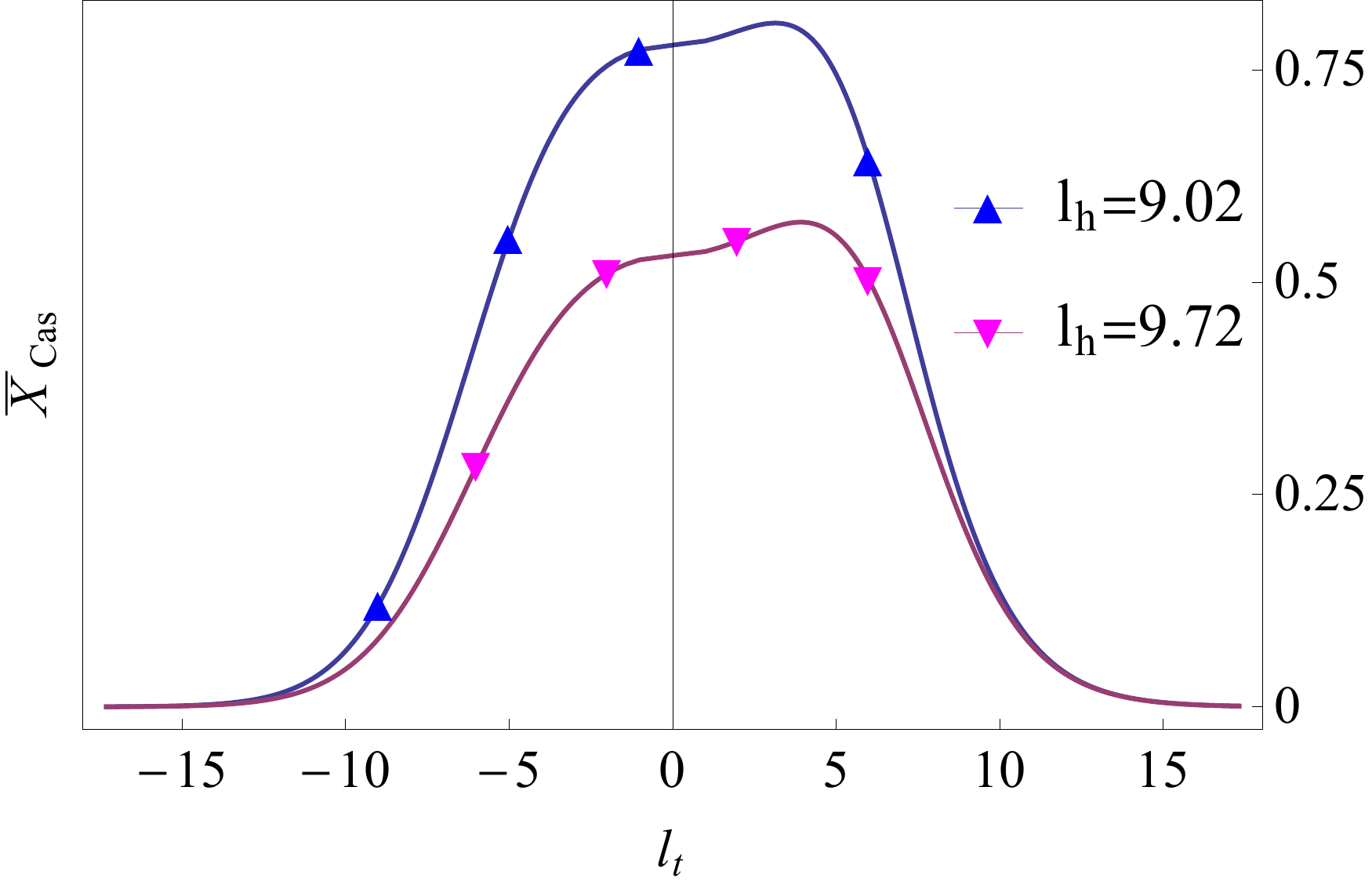}
	\caption{
		The Casimir force as a function of $l_t$ at few values of $l_h$.} 
	\label{fig:dif_lh_as_lt}
\end{figure}
Obviously, for any fixed values of $l_t$ and $l_h$ this is an equation for the scaling function of the pressure in the finite system $p$, i.e., this equation {\it implicitly} determines $p(l_t,l_h)$. 

Since $P[X] \geq 0$ for any $X\in(-\infty,\infty)$ one, setting $X=0$, obtains that $p \geq 0$. The last, of course, shall be expected on physical grounds since it guarantees the stability of the finite film with a given thickness. Since $P[X]$ can not be negative it does not have real roots different from each other - otherwise the derivative will change sign when $X$ passes through such a root. If the real roots are degenerated, however, the  integral in \eq{eq:main_eq_2} diverges and \eq{eq:main_eq} can not have a solution. Thus, all the roots of P[X] are not real and, since $P[X]$ is real, they are conjugated in pairs.  Furthermore, if $x_1=\bar{x}_2$, and $x_3=\bar{x}_4$ are the roots of $P[X]$, it is easy to show that $|x_1|^2 |x_3|^2=p(l_t,l_h)$ and that $x_1+x_2+x_3+x_4=0$. The above implies that $p>0$, i.e., the pressure is strictly positive, and that the general form of the roots is $x_1=a+ I b, x_2=a-I b, x_3=-a+I c, x_4=-a-I c$. Because the roots can not be real and due the corresponding $a\to -a$, $b\to -b$ and $c\to -c$ symmetry in the roots we can take, without lost of generality, $b,c>0$ and $a\ge 0$. Summarizing the above, we come to the conclusion that the general form of the function $f$ is
	\begin{equation}
	\label{eq:Ponx}
	f=\int_{-\infty}^{\infty} \frac{dX}{\sqrt{\left[\left(X-a\right)^2+b^2\right]\left[\left(X+a\right)^2+c^2\right]}},
	\end{equation}
	with
	\begin{equation}
	\label{eq:p_abc}
	p(l_t,l_h)=(a^2+b^2)(a^2+c^2).
	\end{equation}
	This integral can be expressed in terms of elliptic functions \cite{VDD2017,remark}. The result is
	\begin{equation}
	\label{eq:new_integral}
	 f =\frac{2K\left[\sqrt{-\frac{4a^2+(b-c)^2}{4bc}}\right]}{\sqrt{bc}} = \frac{4 K\left(\sqrt{\frac{4 a^2+(b-c)^2}{4 a^2+(b+c)^2}}\right)}{\sqrt{4 a^2+(b+c)^2}},
	\end{equation}
	where $K(k)$ is the complete elliptic integral of the first kind with modulus $k$. It is easy to show that  $b,c$ and $p$ can be expressed in terms of $a,l_t$ and $l_h$. Provided $a\ne 0$, one has
	\begin{equation}
	\label{eq:c}
	c^2=a^2+\frac{1}{2} \left(\sign(l_t)  l_t^2+\frac{1}{3 \sqrt{6}}\frac{l_h^3}{ a}\right)
	\end{equation}
	and 
	\begin{equation}
	\label{eq:b}
	b^2=a^2+\frac{1}{2} \left(\sign(l_t)  l_t^2-\frac{1}{3 \sqrt{6}}\frac{l_h^3}{ a}\right).
	\end{equation}
	Thus, for $p$ one obtains
	\begin{equation}
	\label{eq:p_final_gen}
	p=\left(2 a^2+\frac{1}{2} \sign\left(l_t\right) l_t^2 \right){}^2-\frac{l_h^6}{216 a^2},
	\end{equation}
	where $a$ is to be determined from the equation 
	\begin{equation}
	\label{eq:main}
	1=f(l_t,l_h,a)
	\end{equation}
	as a function of both $l_t$ and $l_h$.The above implies that we have parametrized the behaviour of the pressure in the finite system, as a function of $l_t$ and $l_h$, in terms of a single parameter $a$, which value has to be found from  	\eq{eq:main} for fixed values of $l_t$ and $l_h$.

	We will use the procedure described above in order to determine analytically, see below, the asymptotic behaviour of $X_{\rm Cas}(l_t=0,l_h)$ for $l_h\gg 1$. Before doing that, we will demonstrate that this procedure can be  essentially simplified even further into a new one which does not require any equation to be solved.
	
	The explicit form of \eq{eq:main} is
	\begin{equation}
	\label{eq:gen_main}
	1=\frac{2 K\left(\sqrt{\frac{1}{2}-\frac{6 a^2+\sign\left(l_t\right) l_t^2}{4\sqrt{\frac{1}{4} \left[2 a^2+\sign\left(l_t\right)
				l_t^2\right]^2-\frac{l_h^6}{216 a^2}}}}\right)}{\sqrt[4]{\frac{1}{4} \left[2 a^2+	\sign\left(l_t\right) l_t^2
		\right]^2-\frac{l_h^6}{216 a^2}}}.
	\end{equation}
	
	For any $a\ne 0$, i.e., $a>0$, the solution of \eq{eq:gen_main} can be trivialized by introducing the variables 
	\begin{equation}
	\label{eq:hat_var}
    l_t=a	\hat{l}_t\qquad \mbox{and} \qquad l_h=a \hat{l}_h.
	\end{equation}
	Then, \eq{eq:gen_main} becomes
	\begin{equation}
	\label{eq:gen_main_hat}
	a=\frac{2 K\left(\sqrt{\frac{1}{2}-\frac{6 +\sign\left(\hat{l}_t\right) \hat{l}_t^2}{4\sqrt{\frac{1}{4} \left[2 +\sign\left(\hat{l}_t\right)
					\hat{l}_t^2\right]^2-\frac{\hat{l}_h^6}{216 }}}}\right)}{\sqrt[4]{\frac{1}{4} \left[2+	\sign\left(\hat{l}_t\right) \hat{l}_t^2
			\right]^2-\frac{\hat{l}_h^6}{216}}}.
	\end{equation}
	Thus, for any fixed values of $\hat{l}_t$ and $\hat{l}_h$ by simply evaluating the r.h.s. of \eq{eq:gen_main_hat} one determines $a$ and, from 	\eq{eq:hat_var}, the values of $l_t$ and $l_h$. Plugging these values in 	\eq{eq:p_final_gen} one directly arrives at $p=p(l_t,l_h)$. 
	
	Let us note that the case $a=0$ leads to $l_h=0$, a case that has been studied earlier \cite{K97}. Within our approach, however, we can get expressions much simpler than the ones reported in \cite{K97}.  Indeed, when $a=0$ one has
	\begin{equation}
	\label{eq:h0_def}
	p=b^2c^2 \qquad \mbox{and} \qquad \sign(l_t)l_t^2=b^2+c^2.
	\end{equation}
	As it immediately follows form \eq{eq:h0_def}, such a representation is only possible for $\tau \ge 0$. 
	Then, from \eq{eq:new_integral} one obtains
	\begin{equation}
	\label{eq:h0_p}
	1=2 K\left(\sqrt{\frac{1}{2}-\frac{l_t^2}{4
			\sqrt{p}}}\,\right){\Bigg /}{\sqrt[4]{p}}, \qquad \tau\ge 0,
	\end{equation}
	solving which one determines the pressure $p(l_t,0)$ in the finite system. Despite this equation is simpler than the set of expressions for $p$ derived in \cite{K97}, its solution can be trivialized by introducing the variable $\tilde{l}_t=l_t/ \sqrt[4]{p}$. Then, 	\eq{eq:h0_p} becomes
	\begin{equation}
	\label{eq:h0_p_final}
	l_t=2 \tilde{l}_t K\left(\sqrt{\frac{1}{2}-\frac{ \tilde{l}_t^2}{4}}\,\right), \qquad \tau\ge 0.
	\end{equation}
	Now the procedure is obvious: for a fixed $\tilde{l}_t$ one calculates $l_t$, wherefrom one immediately finds $p=(l_t/\tilde{l}_t)^4$. We recall that when $h=0$ and $\tau \ge 0$, one has $p=X_{\rm Cas}(l_t\ge 0,l_h=0)$.  The above can be easily extended to cover also the case $l_t<0, l_h=0$. To do that it is enough to check that when $b=c$ one has $l_h=0$ and that then $\sqrt{p}=a^2+b^2, \frac{1}{2} \sign(l_t)l_t^2=a^2-b^2$, i.e.,  one can have both positive and negative values for $l_t$. Next, from the second of the representations reported in  	\eq{eq:new_integral}, one derives that 
\begin{equation}
	\label{eq:h0_p_gen}
	1=2 K\left(\sqrt{\frac{1}{2}-\frac{\sign(l_t) l_t^2}{4
			\sqrt{p}}}\,\right){\Bigg /}{\sqrt[4]{p}},
\end{equation}
which is an obvious generalization of \eq{eq:h0_p} that reduces to it for $\tau>0$. Thus, when $\tau<0,h=0$, one can proceed in the same way as outlined above for $t>0,h=0$, by just taking in addition into account that then $p_b=l_t^4/2$. 
	
Using the exact expressions for the scaling function of the Casimir force one can derive the corresponding asymptotes for large $l_t$, or large $l_h$ values.

The asymptotes of $X_{\rm Cas}(l_t,l_h=0)$, with $|l_t|\gg 1$, can be found from the explicit analytical expressions given in \eq{CasimiralphaeqpiKrech}. In terms of $l_t$ one obtains
\begin{equation}
\label{eq:as_h0}
X_{\rm Cas}(l_t,l_h=0)\simeq \left\{ \begin{array}{cc} 
16\; l_t^4 \; \exp[-l_t], & l_t>0 \\
16 \; l_t^4 \; \exp[l_t/\sqrt{2}], & l_t<0.
\end{array}\right .
\end{equation}
The above expressions are equivalent with the corresponding results derived in \cite{K97} in terms of other variables. The corresponding result for $T<T_c$ contains the same exponential decay, as predicted in \cite{PE92}.

The asymptotes $X_{\rm Cas}(l_t=0,l_h)$, with $|l_h|\gg 1$, as far as we are aware of, have never been reported before. We start from equation $1=f(l_t=0,l_h,a)$ where $f$ is given by \eq{eq:new_integral} with $l_t$=0. It is easy to show that this equation then becomes
\begin{equation}
\label{eq:as_lh}
l_h=\frac{2 \sqrt{6} \sqrt[6]{1-\varepsilon} K\left(\sqrt{\frac{1}{2}-\frac{3}{2 \sqrt{\varepsilon }}}\right)}{\sqrt[4]{\varepsilon }}
\end{equation}
with
\begin{equation}
\label{eq:epsilon}
\varepsilon=1-\frac{l_h^6}{216 a^6}
\end{equation}
and 
\begin{equation}
\label{eq:p_Y}
p=\frac{1}{36}\frac{(\varepsilon +3)}{(1-\varepsilon )^{2/3}} \;  l_h^4.
\end{equation}
It is trivial to show that when $l_t=0$ one has 
\begin{equation}
\label{eq:pb_ltis0}
p_b=\frac{1}{12} l_h^4.
\end{equation}
Equations \eref{eq:as_lh}, \eref{eq:p_Y} and \eref{eq:pb_ltis0} provide the equivalence to the case of $l_t=0, l_h\ne 0$ of the results reported in \eq{CasimiralphaeqpiKrech} for the case $l_t\ne 0, l_h=0$. Indeed, from  \eq{eq:p_Y} and \eq{eq:pb_ltis0} one obtains
\begin{equation}
\label{eq:Xcaslt0}
X_{\rm Cas}(l_t=0,l_h)=\frac{1}{12} l_h^4 \left[\frac{1}{3}\frac{(\varepsilon +3)}{(1-\varepsilon )^{2/3}} -1\right].
\end{equation}
This equation can be read in two different ways. First, for a given value of $l_h$ one determines $\varepsilon=\varepsilon(l_h)$ from  \eq{eq:as_lh}, wherefrom one also calculates $X_{\rm Cas}(l_t=0,l_h)$, using \eq{eq:Xcaslt0}. Second, for a given fixed value of $\varepsilon$   from  \eq{eq:as_lh} one determines $\l_h$ by simple evaluation of the r.h.s. of this expression, wherefrom one calculates  $X_{\rm Cas}(l_t=0,l_h)$  from \eq{eq:Xcaslt0}. 

It is easy to show that the r.h.s. of \eq{eq:as_lh} diverges only when $\varepsilon\to 0^+$. Then, after expanding about $\varepsilon=0$, one obtains that   \eq{eq:as_lh} becomes
\begin{equation}
\label{eq:lh_Y_about_1}
l_h = \ln \left(\frac{576}{\varepsilon }\right)+\frac{7}{48} \varepsilon  \ln (\varepsilon ) +\cal{O}(\varepsilon). 
\end{equation}
Solving this equation about $\varepsilon$ and placing the result  in \eq{eq:Xcaslt0}, one, keeping only the leading order term, arrives at
\begin{equation}
\label{eq:as_XCas_h}
p-p_b\equiv X_{\rm Cas}\simeq 48\; l_h^4\;\exp(- |l_h|), \qquad \mbox{when} \qquad l_t=0, |l_h|\gg 1.
\end{equation}
Thus, the Casimir force decays exponentially not only for large values of $|l_t|$, but also for large values of $|l_h|$.
	
After showing how one can easily obtain precise numerical results within the considered model, we pass to derivation of some facts for the scaling function of the force and to present some results out such   numerical calculations. Let us note that some of these properties we are going to obtain are known from before based on some qualitative or semi-quantitative arguments. In our treatment we will simply obtain them as mathematical facts following from the above expressions. 
	
The function $f$ defined by \eq{eq:main_eq} is a monotonically decreasing function of $p$. Thus, if it exists, there is only one solution of \eq{eq:main_eq} for any fixed pair of values $(l_t, l_h)$. 

Setting in $P[X]$ the bulk value of the order parameter, i.e., $X=X_b$, and taking into account \eq{pbX1} one obtains that $P[X_b]=p(l_t, l_h)-p_b(l_t, l_h)>0$. Then, taking into account \eq{CasimirX1}, one concludes that $X_{\rm Cas}(l_t,l_h)>0$, i.e., the Casimir force is {\it repulsive} under $(+,-)$ boundary conditions for any values of $l_t$ and $l_h$. 
	
	Taking into account the symmetry of the equation under the change of the variables $(X\to -X, l_h\to -l_h)$, one immediately obtains that $p(l_t,l_h)=p(l_t,-l_h)$. This property is also obvious from 	Eqs. \eref{eq:p_final_gen} and \eref{eq:gen_main}. It is easy to check that the same property remains valid also for $p_b$. Thus, one concludes that 
	\begin{equation}
	\label{eq:symmetry}
	X_{\rm Cas}(l_t,l_h)=X_{\rm Cas}(l_t,-l_h).
	\end{equation}
	
The above property, plus the easily verifiable due to \eq{eq:main_eq} fact that the derivative of $p$ with respect to $l_h$ can be zero only for $l_h=0$, immediately leads to the conclusion that $X_{\rm Cas}(l_t,l_h)$ has an extremum for $l_h=0$. As it follows from  \eq{CasimiralphaeqpiKrech}, it is achieved for $k_{\rm max}^2=0.826115$ determined as the root of the equation  $2=K(k)/E(k)$, and the extremum is in fact a maximum. For $k=k_{\rm max}$ one obtains that the normalized force is $5.64$, that is attained at $l_t=-5.30$. Thus, in the $(l_t,l_h)$ plane the force has a {\it single global maximum}, i.e.
	\begin{equation}
	\label{eq:max}
	\max_{(l_t,l_h)}	X_{\rm Cas}(l_t,l_h)=	X_{\rm Cas}(l_t=-5.30,l_h=0). 
	\end{equation}

In the figures that follow we illustrate the above properties of the scaling function of the Casimir force under $(+,-)$ boundary conditions.

Figure \ref{fig:dif_lh_as_lt} shows the behaviour of the Casimir force as a function of $l_t$ at few values of $l_h$. Let us note that the results are normalized with the Casimir amplitude $\Delta_{\rm Cas}^{(+,+)}$, see \eq{deltarelpppm}, which purpose is two-fold: {\it i)} In this way one gets rid of the dependence on the nonuniversal factor $g$ - see  \eq{CasimirX}, and {\it ii)} Facilitates its comparison with the numerically obtained results for the behaviour of the force as a function of $l_h$ for the case $T=T_c$ which have been reported in Ref \cite{VD2013}. On wish, as \eq{deltarelpppm} shows, it is trivial to normalize the data to $\Delta_{\rm Cas}^{(+,-)}$, instead of to $\Delta_{\rm Cas}^{(+,+)}$. The normalized  scaling function of the force is denoted by $\bar{X}_{\rm Cas}(l_t,l_h)$, where 
\begin{equation}
\label{eq:Xbar}
\bar{X}_{\rm Cas}(l_t,l_h)\equiv X_{\rm Cas}(l_t,l_h)/\Delta_{\rm Cas}^{(+,+)}. 
\end{equation}

\begin{figure}[htb!]
	\centering
	\includegraphics[width=\columnwidth]{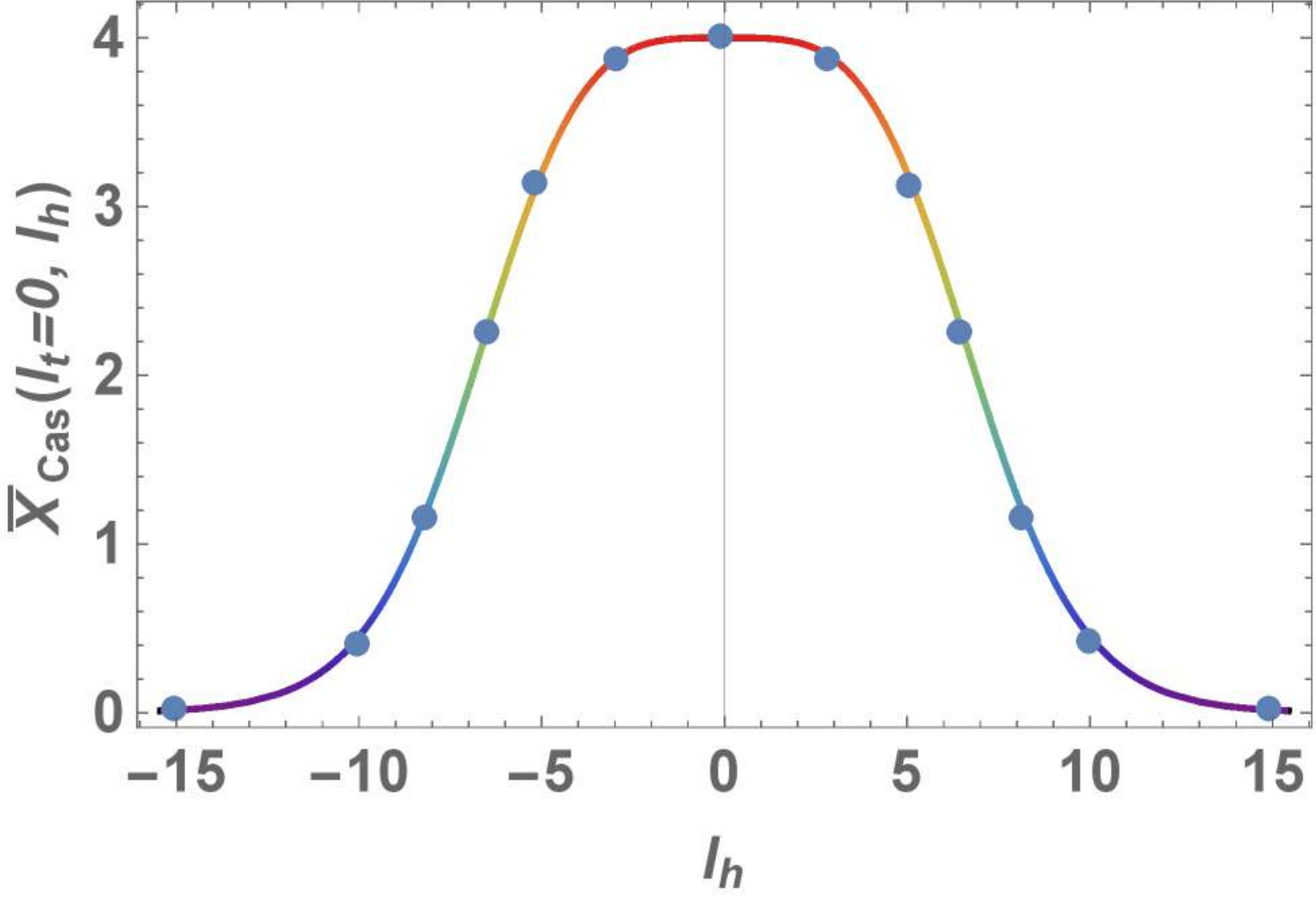}
	\caption{
		A comparison of the behaviour of the scaling function of the Casimir force derived analytically, solid line, with that one obtained numerically in \cite{VD2013}, the points. Following \cite{VD2013}, the scaling function $X_{\rm Cas}(l_t,l_h)$ is normalized per $\Delta_{\rm Cas}^{(+,+)}$ , i.e., the figure shows the behaviour of $\bar{X}_{\rm Cas}(l_t,l_h)=X_{\rm Cas}(l_t,l_h)/\Delta_{\rm Cas}^{(+,+)}$ as a function of $l_h$ for $l_t=0$. } 
	\label{fig:comparison}
\end{figure} 
With respect of the force one observes that, as expected, the field strongly influences the behaviour of the force. As it has been already proven above, the strongest force is achieved at $l_h=0$. For moderate values of the field the maximal value of the force is at {\it negative} values of $l_t$ - see Fig. \ref{fig:dif_lh_as_lt}, left, while for stronger fields the maximum value is at {\it positive} values of $l_t$ - see Fig. \ref{fig:dif_lh_as_lt}, right. This happens because of the competing effects due to the temperature and the field on the fluctuations of the system in different regions of the $(T,h)$ plane.

The comparison of our analytically derived results for the field dependence of the force with those numerically obtained in \cite{VD2013} for the case of $T=T_c$ is presented in Fig. \ref{fig:comparison}. One observes an excellent agreement between them.

\begin{figure}[h!]
	\includegraphics[width=8cm]{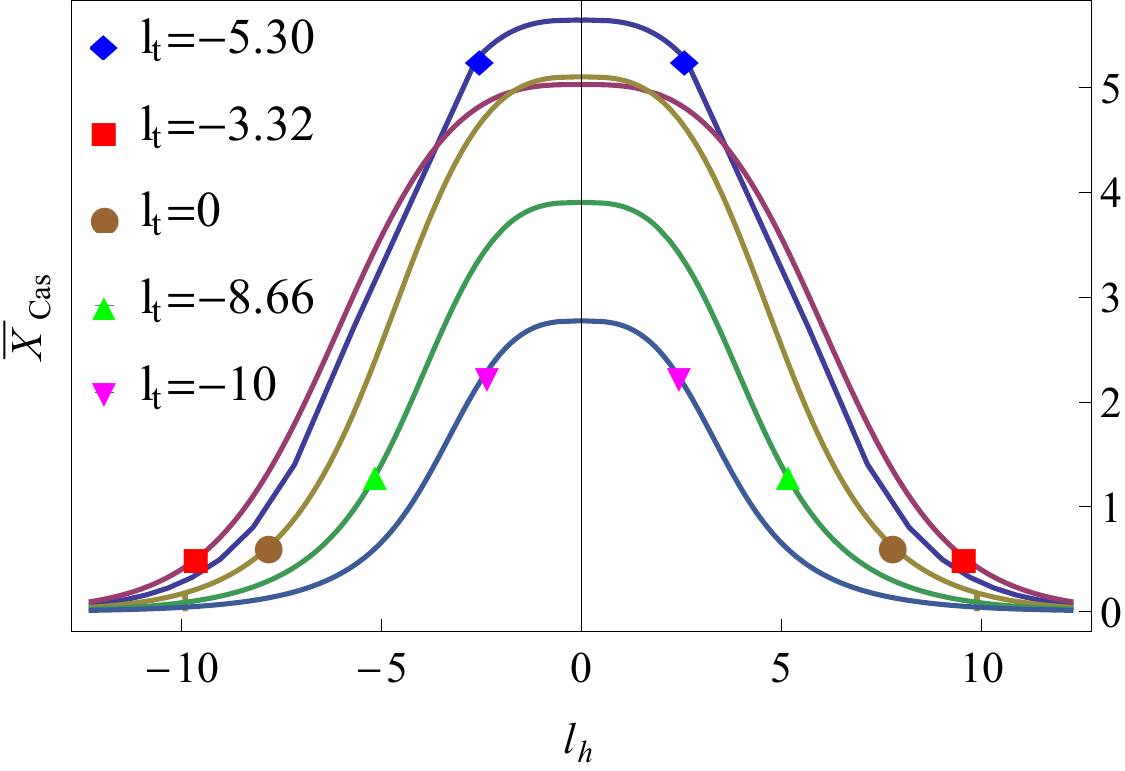}
	\includegraphics[width=8cm]{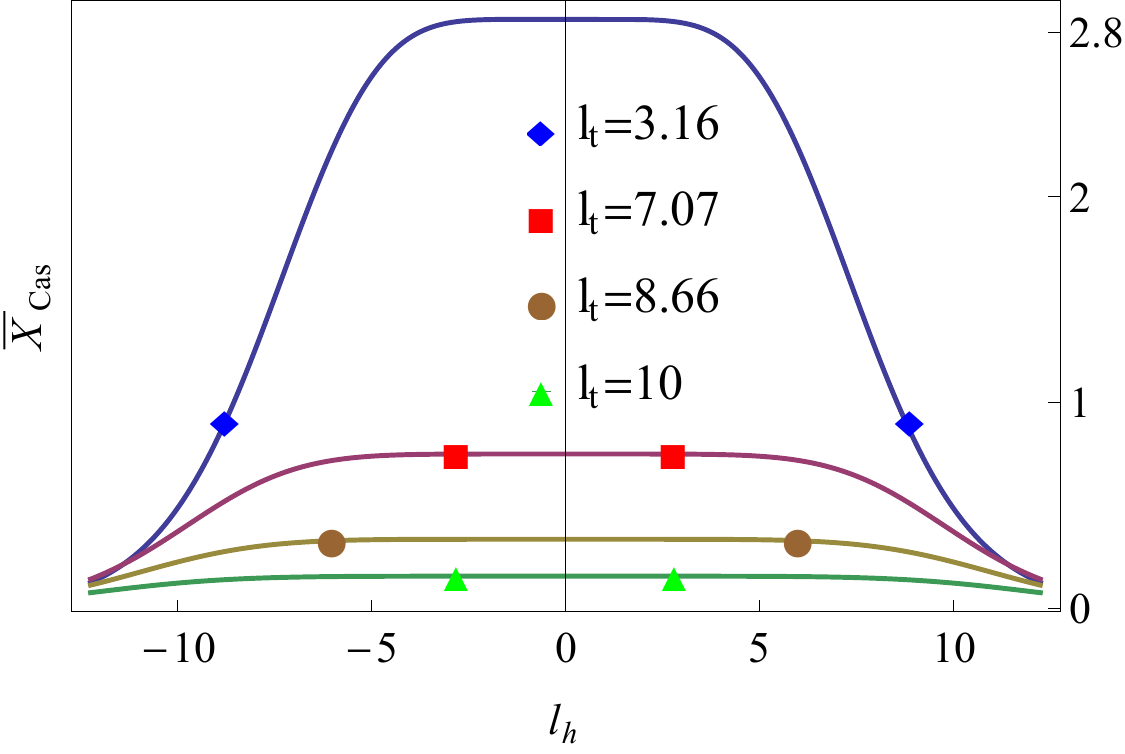}
	\caption{
		The Casimir force as a function of $l_h$ at few values of $l_t$.} 
	\label{fig:dif_lt_as_lh1}
\end{figure}

Figure \ref{fig:dif_lt_as_lh1} shows the behaviour of the Casimir force as a function of $l_h$ at few values of $l_t$. The results are again normalized with the Casimir amplitude $\Delta_{\rm Cas}^{(+,+)}$. 

With respect of the force one observes the following: 
\begin{itemize}
	\item For any fixed $l_t$ the force, as analytically proven above - see \eq{eq:max}, is a symmetric function of $l_h$. 
	\item The strongest force is observed for $l_t=-5.30$.  The behaviour of the force for this particular value of $l_t$ as a function of $l_h$ is also shown.  
	\item The maxima of the force, for values of $l_h$ at $O(1)$ distance from each other, are quite close to each other. The last implies that in the surface $(l_t,l_h)$ of the force there will be plateau-like regions for intervals of $l_h$ in which the force stays almost unchanged. 
	\item As it is easy to be demonstrated from \eq{eq:main_eq} by differentiating the both sides of that equation with respect to $l_h$, the derivative of the pressure $\partial p/ \partial l_h$ in the finite system is proportional to $l_h^2$ near $l_h=0$. Similar is also true for the behaviour of the bulk pressure $p_b$. These  facts  explain why the Casimir force as a function of $l_h$ has a plateau-like region near $l_h=0$. 
\end{itemize}

\begin{figure}[h!]
	\includegraphics[width=\columnwidth]{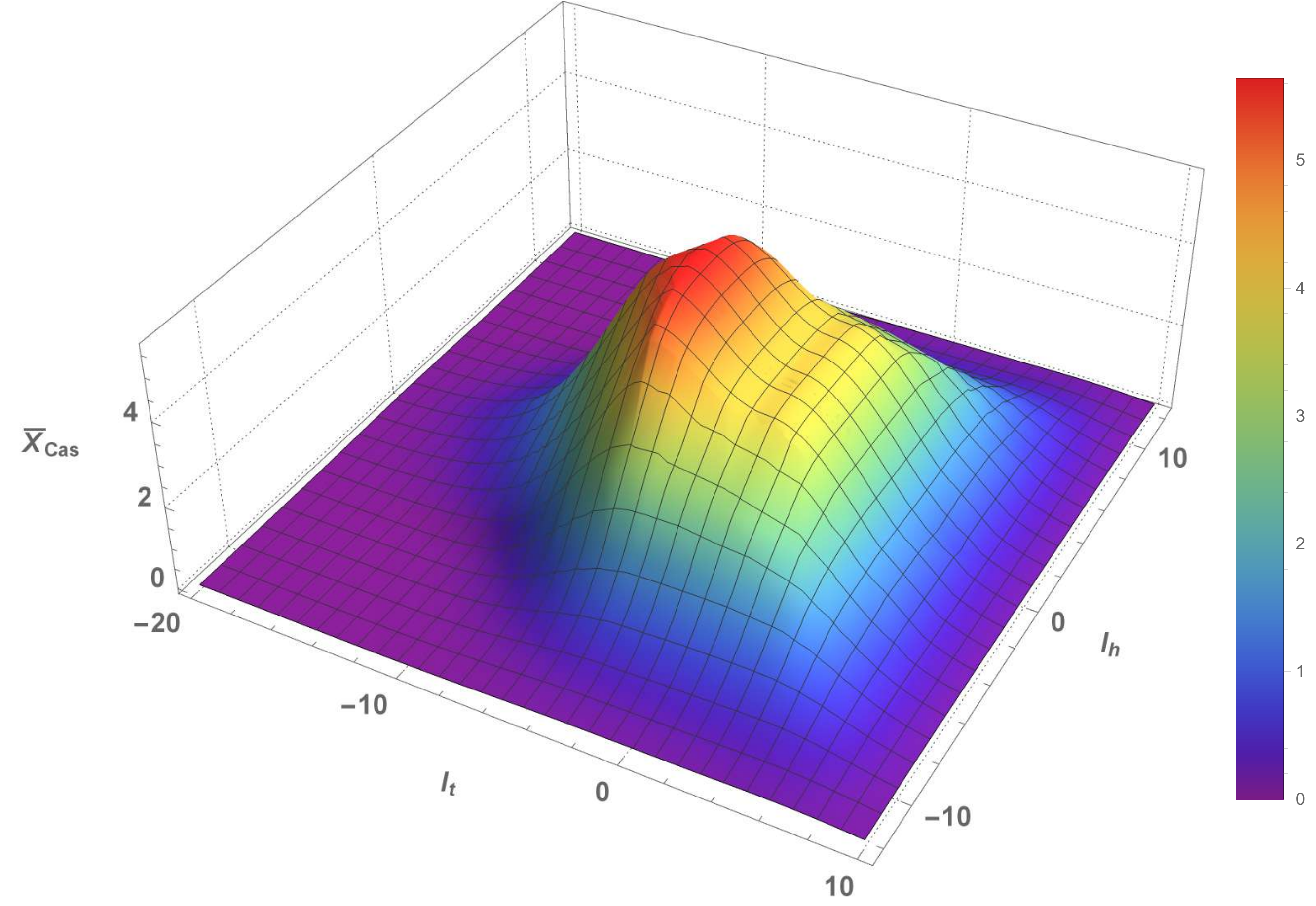}
	\caption{
		The Casimir force as a function of {\it both} temperature $l_t$ and field $l_h$ scaling variables.}
	\label{fig:CF_all}
\end{figure}

The overall (temperature-field) relief map of the force is shown in Fig. \reff{fig:CF_all}. The global maximum of the force, attained at $l_t=-5.30, l_h=0$, is clearly visible. The value of the maximum after the normalization given in  \eq{eq:Xbar} is
\begin{equation}
\label{eq:max_CF}
\max_{\left(l_t, l_h\right)}\bar{X}_{\rm Cas}(l_t,l_h)=\frac{X_{\rm Cas}(l_t=-5.30,l_h=0)}{\Delta_{\rm Cas}^{(+,+)}}=5.64. 
\end{equation} 
One can check - one by one - that all the other properties of the force listed above, are obeyed. 

\section{DISCUSSION AND CONCLUDING REMARKS} 

In the current article we have obtained analytical results for the temperature and ordering field behaviour of the Casimir force, see Eqs. \reff{CasimirX} and \reff{CasimirX1}, in one widely used model - the Ginsburg-Landau mean field model. More precisely, we have determined the scaling function $X_{\rm Cas}(l_t,l_h)$ of the force - see Fig. \reff{fig:CF_all}. In \eq{CasimirX1}, the quantity $p(l_t,l_h)$ is the solution of \eq{eq:main_eq}, while $p_b(l_t,l_h)$ is obtained from the cubic equation of the bulk order parameter profile for which $P_b$ has its maximum. The simplest procedure, however, how one can determine $p(l_t,l_h)$ by simple evaluation of some expressions without the need to solve whatever equations,  is given by Eqs. 	\eref{eq:hat_var} and 
\eref{eq:gen_main_hat} and explained in the text around them.

We focused on a model with a film geometry under the so-called $(+,-)$ boundary conditions which correspond to physical situation in which the confining surfaces have  strongly adsorbing but competing properties. For the case of a simple fluid this is the situation of one wall  is strongly preferring the liquid, i.e., it is wet by the fluid, while the other one favours the vapor phase, i.e., one observes drying of that surface. In the case of a binary liquid mixture the $(+,-)$ boundary conditions correspond to a situation in which one of the walls picks up one of the components while the other surface selects the other one. The degree of preference of the corresponding boundary is normally modeled by applying a proper surface field acting on that surface.  In our case these boundary fields have been set to $\pm \infty$ so that at the boundaries the order parameter  behaves in accord with \eq{eq:pl_min_bc}. Let us note that the model we considered can be obtained as a special case of a more general model \cite{PE92,SOI91}, see also Refs. \cite{PE90,PEN91}, that allows for finite surface fields and surface enhancements. Then, the model enjoys a rich phase diagram that has been obtained in \cite{SOI91} and \cite{PE92}. The model we are studying in the current article can be considered as a particular limiting case of that more general model with the surface fields amplitudes approaching plus infinity on one surface and minus infinity on the other. Let us note that under such a limit the wetting temperature of the corresponding semi-infinite system $T_w$ diminishes. 

We have proven several properties of the scaling function $X_{\rm Cas}(l_t,l_h)$. Further more, we have visualized its behaviour as a function of $l_t$ at few values of $l_h$ in Fig. 	\ref{fig:dif_lh_as_lt}, and as a function of $l_h$ at few values of $l_t$ in Fig. \ref{fig:dif_lt_as_lh1}, respectively. Needless to say, for $l_h=0$ our results reduce to the ones previously obtained in \cite{K97}. It is worth emphasizing, that in our approach we derived the behaviour of the force without using the detailed properties for the behaviour of the order parameter profile, as it has been done in the approach used in \cite{K97} for $(+,-)$ boundary conditions, in \cite{K97,DRB2009,DVD2015,DVD2016} for $(+,+)$ and in \cite{GaD2006} for Dirichlet-Dirichlet boundary conditions, respectively. Finally, a comparison of the behaviour of the scaling function of the Casimir force derived analytically with that one obtained numerically in \cite{VD2013} is shown in Fig. \ref{fig:comparison}. One observes excellent agreement between them.  
\begin{figure}[h!]
	\includegraphics[width=0.5\columnwidth]{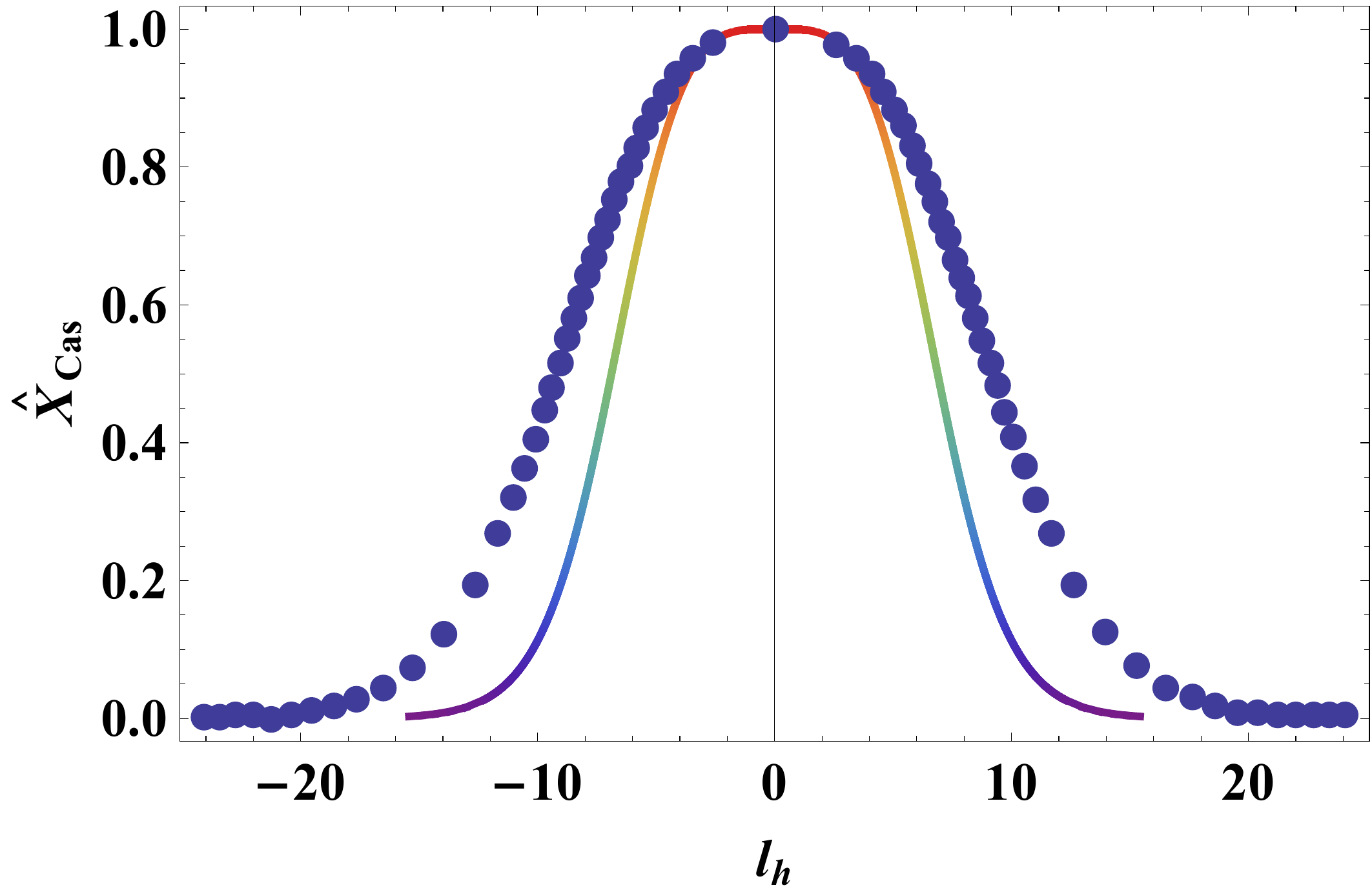}
	\includegraphics[width=0.5\columnwidth]{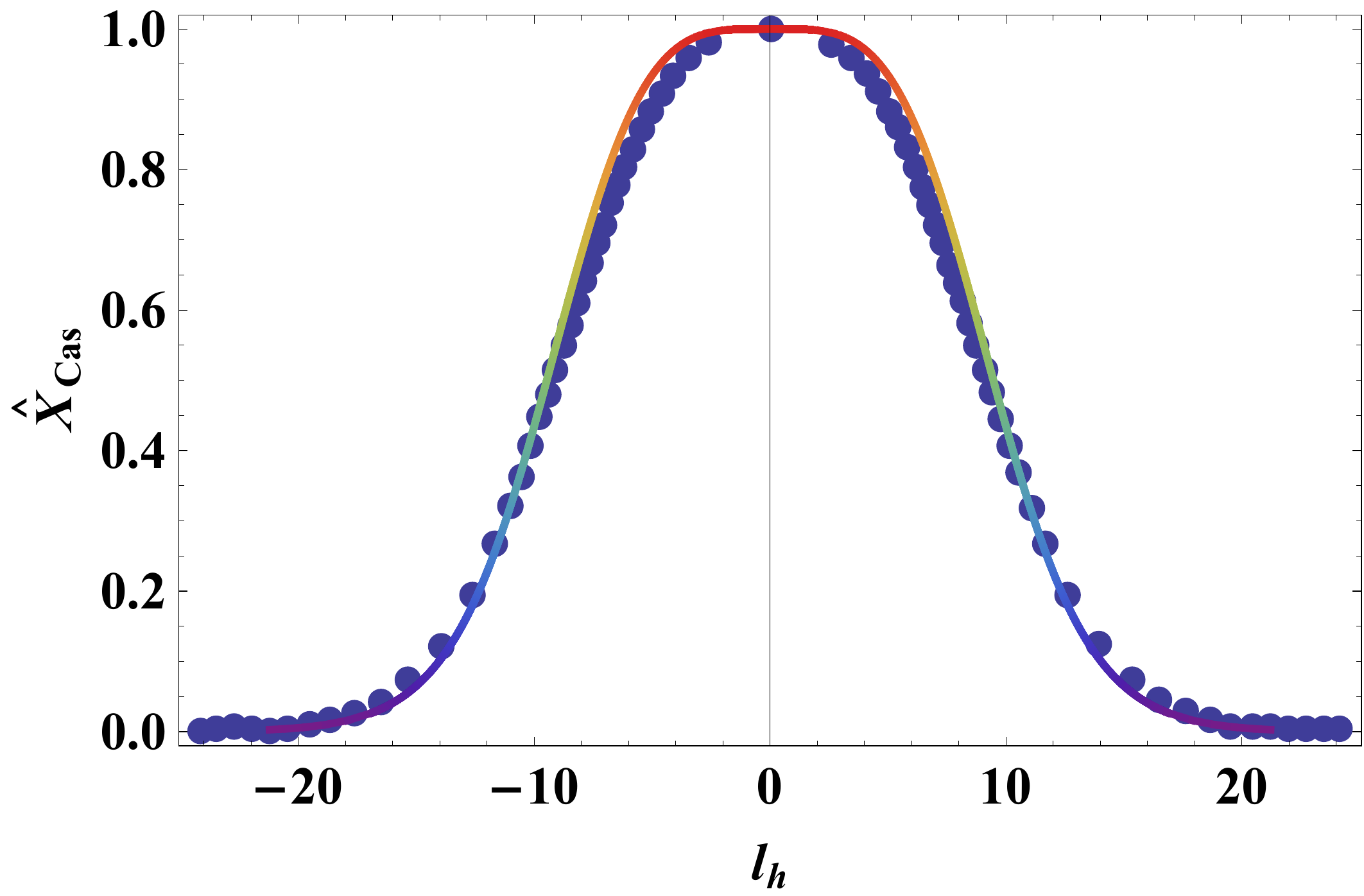}
	\caption{A comparison of the scaling function of the 3d Ising model obtained with a Monte Carlo simulation (dots) to the mean-field result derived analytically here (curves). Both results are scaled vertically with the value at the corresponding Casimir amplitude, that is $\hat{X}_{\rm Cas}(l_t,l_h) = X_{\rm Cas}(l_t,l_h)/X_{\rm Cas}(0,0)$ is plotted. In that way there is no dependence on the non-universal quantity $g$ - see \eq{CasimirX}. The right figure represents the data when additional scaling along the horizontal axis of the mean field data with a factor 1.375 is performed. One observes a perfect overlapping of the both curves.}
	\label{fig:MF_MC}
\end{figure} 

Based on the derived analytical results, we confirm that in the finite system there is no phase transition of its own at any finite values of $l_t$ and $l_h$ thus verifying the general theoretical statements that such a phase transition can be observed only below the wetting temperature $T_w$ of the corresponding semi-infinite system \cite{PEN91,PE90,BLF95,BELF96,BLF95b,BLM2003,BHVV2008,AB2009,VD2013}. We recall that, as a rule, $T_{w}$ is at a finite distance below $T_c$ and its value depends on the value of the boundary field in the semi-infinite system, say $h_1$, i.e., $T_w=T_w(h_1)$.  Let us note that the lack of phase transition of its own in the finite system close to the critical point of the infinite system $T_c$ \cite{BDT2000,P90,Ba83} for $(+,-)$ boundary conditions is in a striking contrast with the symmetric case, for which the finite system does posses a phase transition of its own at some $T_{c,L}$ that is algebraically close to $T_c$ on the scale of $L^{-1/\nu}$ where it undergoes the capillary condensation phase transition. For the model considered here, the case of $(+,+)$ boundary conditions has been studied in details in \cite{DVD2015,DVD2016}. The difference between the two cases is, however, not so difficult to be understood on a general reasoning. While under $(+,+)$ boundary conditions the $"+"$ phase is favoured by the boundary conditions and the small negative
field bolsters the appearance of the competitive $"-"$ phase, under $(+,-)$ boundary conditions the both competing phases are present in the system for any moderate values of the thermodynamic parameters. Therefore, no spontaneous symmetry breaking can occur in the system - the phase transition occurring in the bulk at a temperature $T_c$ is suppressed, and instead one observes the formation of an interface between coexisting phases stabilized by the surface fields. In order this to be overcome, one needs much smaller surface fields to be acting on the bounding surfaces of the system. For such fields one can have a layer of a, say, $"+"$ phase near the border that from a microscopic becomes macroscopically thick, i.e., a finite portion of the film thickness. Such phenomena is known as wetting phenomena, if one performs the limit of the semi-infinite system. In a finite system one will have an interface initially positioned near one of the borders that moves inside of the system. One normally terms this last phenomena attachment - detachment phase transition \cite{BLF95,BELF96}. Obviously, both the phase behaviour of such a system, as well as the behaviour of the Casimir force in it, will be much richer than in the situation we have studied. Let us note, that when the surface field  $h_1$ is such that critical wetting occurs at a single wall ($L = \infty$) at a transition temperature $T_w(h_1)$, the force is predicted \cite{PE92} to change from repulsive to attractive (at large $L$) when the temperature is reduced below $T_w(h_1)$.  Furthermore, when $T_c > T> T_w$ the system is in a single soft mode phase that is characterized, for $h=0$, by a $(+,-)$ interface located at the centre of the film, a transverse correlation length $\xi_{\|} \sim  \exp(L)$, and, as we already know, a Casimir force that is repulsive.

The only other exact results that exist in the case of asymmetric boundary conditions pertain to the two-dimensional Ising model \cite{ES94,NN2008,NN2009}. In \cite{NN2008,NN2009} the authors consider the case of $h_1=h_2=O(1)$ and $h_1=-h_2=O(1)$. The exact results obtained in \cite{NN2008,NN2009} demonstrate that indeed the force changes sign below $T_w$. It is shown that the overall behaviour of the force depends, as expected, on the values of the surface fields. For $(+,-)$ boundary conditions the force behaves similarly to what has been demonstrated in the current article. It shall be, however, mentioned that it decays with $T$ much slower to zero than in the mean-field case. Indeed, within the $d=2$ Ising model when an interface is enforced by the boundary conditions the excess free energy decays as $L^{-3}$ for $T_w<T<T_c$ 
\cite{ES94,NN2008,NN2009,RZSA2010,AM2010} and exponentially below $T_w$.  
In the more general case when 
an interface exists the arguments stated in Refs. \cite{PEN91,PE92}
about the form of the {\it singular} contributions to the excess free energy
due to the interface wandering suggest that: when $d<3$ and $T_w <T \ll T_c$,
then $F_{\rm Cas}\sim L^{-\tau-1}$, where $\tau=2(d-1)/(3-d)$ is the
so-called exponent for thermal wandering that has been introduced in Refs.  \cite{LF86,LF86b} in order to account phenomenologically	for fluctuation effects at wetting transitions. Let us note that for the proper derivations of the size effects in the case of an existing interface in the system one has to take into account the effects due to the capillary fluctuations of the interface. When $d\geq 3$ the situation seems simpler and one again expects exponential decay of the force, just as in the case with no interface in the
system. The explicit mean field results \cite{PE92,K97}, as well as our results, suggest $F_{\rm Cas}
(T,h=0,L)\sim \exp(-L/(2\xi_t))$ and $F_{\rm Cas}
(T=T_c,h,L)\sim \exp(-L/\xi_h)$.

For d=3 Monte Carlo results are available for the case $T=T_c$ \cite{VD2013}. For the case considered here of infinite boundary fields (strong adsorption) the scaling function of the force in our model turns out to be very similar to the corresponding one for $d=3$. Even, after a proper rescaling, one obtains that the both scaling functions almost overlap - see Fig. \ref{fig:MF_MC}. One is tempting to make the hypothesis that this is so since the bulk phase transition is suppressed under the $(+,-)$ boundary conditions and thus, in the presence of a magnetic field, the correlations of the system become Gaussian-like even close to $T_c$.  In any case, if confirmed with additional Monte Carlo data, the above observation can be used to obtain good approximate data for the Casimir force in the 3d Ising model under $(+,-)$ boundary conditions in the presence of an ordering field.

\section{ACKNOWLEDGMENTS}
The authors gratefully acknowledge the  financial support via contract DN 02/8 of Bulgarian NSF. D.D. is thankful for the fruitful discussions with S. Dietrich, A. Maci{\`o}{\l}ek  and O. Vasilyev. 

\smallskip
 
\Large{\bf References} 





\end{document}